\documentclass[a4paper]{article}
\usepackage{Odyssey2024}
\usepackage{epsfig,amssymb,amsmath}
\usepackage{mathtools}
\ninept

\usepackage{amsmath,graphicx,booktabs,amsfonts,amssymb,utfsym,multirow,makecell,url,color, colortbl,subcaption,siunitx}

\title{Additive Margin in Contrastive Self-Supervised Frameworks to Learn Discriminative Speaker Representations}

\name{Theo Lepage, Reda Dehak}

\address{
    EPITA Research Laboratory (LRE), France \\
    {\small \tt theo.lepage@epita.fr, reda.dehak@epita.fr}
}

\begin{document}
\maketitle

\begin{abstract}
Self-Supervised Learning (SSL) frameworks became the standard for learning robust class representations by benefiting from large unlabeled datasets. For Speaker Verification (SV), most SSL systems rely on contrastive-based loss functions. We explore different ways to improve the performance of these techniques by revisiting the NT-Xent contrastive loss. Our main contribution is the definition of the NT-Xent-AM loss and the study of the importance of Additive Margin (AM) in SimCLR and MoCo SSL methods to further separate positive from negative pairs. Despite class collisions, we show that AM enhances the compactness of same-speaker embeddings and reduces the number of false negatives and false positives on SV. Additionally, we demonstrate the effectiveness of the symmetric contrastive loss, which provides more supervision for the SSL task. Implementing these two modifications to SimCLR improves performance and results in 7.85\% EER on VoxCeleb1-O, outperforming other equivalent methods.
\end{abstract}

\section{Introduction}
\label{sec:intro}

\begin{figure*}[t]
    \centering
    \begin{subfigure}[T]{0.5\textwidth}
        \centering
        \includegraphics[width=0.93\linewidth]{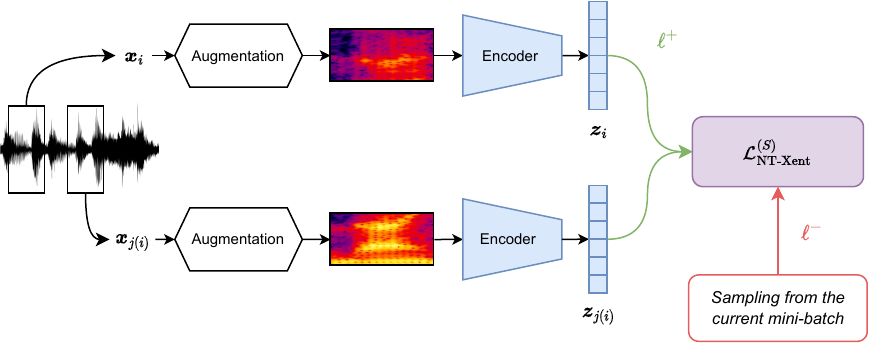}
        \caption{SimCLR}
    \end{subfigure}%
    \begin{subfigure}[T]{0.5\textwidth}
        \centering
        \includegraphics[width=0.93\linewidth]{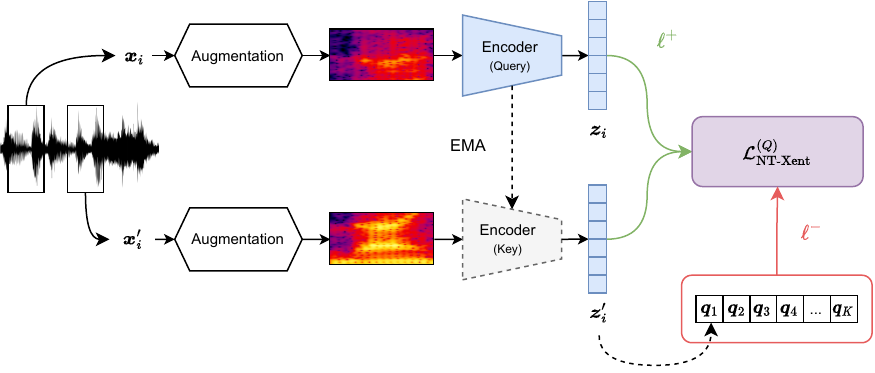}
        \caption{MoCo}
    \end{subfigure}
    \caption{Diagram of our contrastive self-supervised training framework to learn speaker representations.}
    \label{fig:training_framework}
\end{figure*}

\newcommand\blfootnote[1]{%
  \begingroup
  \renewcommand\thefootnote{}\footnote{#1}%
  \addtocounter{footnote}{-1}%
  \endgroup
}

\blfootnote{The code associated with this article is publicly available at\\ \url{https://github.com/theolepage/sslsv}.}

Most Speaker Recognition (SR) methods aim at learning an embedding space making each class distribution compact with a large dispersion between classes. The objective is to learn representations with small intra-speaker distances and large inter-speaker distances while being robust to extrinsic information. Different speech feature extraction methods and machine learning frameworks were proposed for this task. The i-vectors \cite{dehakFrontEndFactorAnalysis2011} is an unsupervised method that learns an embedding capturing the main variabilities of the speech signal and relies on a back-end (LDA and/or PLDA) to focus on the speaker or language information \cite{dehakLanguageRecognitionIvectors2011}. Several solutions based on deep neural networks were recently proposed to learn this embedding space \cite{villalbaStateoftheartSpeakerRecognition2020}. The d-vector \cite{varianiDeepNeuralNetworks2014}, and the x-vectors \cite{snyderXVectorsRobustDNN2018, chungDelvingVoxCelebEnvironment2020, chungDefenceMetricLearning2020} approaches embed speakers audio features into a vector space using deep neural networks.

The drawback of these supervised methods is that they require large labeled datasets during training. Despite the impressive progress made with supervised learning, this paradigm is now considered a bottleneck for building more intelligent systems. Manually annotating data is complex, expensive, and tedious, especially when dealing with signals such as images, text, and speech. Moreover, the risk is creating biased models that do not perform well in real-life scenarios, notably in difficult acoustic conditions.

Recently, methods based on Self-Supervised Learning (SSL) have been developed to train deep neural networks in an unsupervised way by benefiting from massive amounts of unlabeled data. Motivated by the performance obtained by these approaches in computer vision and natural language processing (NLP), several techniques have been applied to learn speaker representations \cite{stafylakis19_interspeech, Cho2020, xiaSelfSupervisedTextIndependentSpeaker2021, zhangContrastiveSelfSupervisedLearning2021, choNonContrastiveSelfsupervisedLearning2022,lepageLabelEfficientSelfSupervisedSpeaker2022}. These frameworks rely on the assumption that segments extracted from the same utterance belong to the same speaker while those from different utterances belong to distinct speakers. This assumption does not always hold (\textit{class collision} issue), but the impact on the training convergence is negligible. The main challenge resulting from this training scenario is that segments extracted from the same utterance share several similar characteristics, such as channel, language, speaker, and sentiment information \cite{9893562}. Thus, augmenting speech segments is fundamental to only encode speaker-related information.

Most SSL methods proposed for Speaker Verification (SV) are contrastive. SimCLR \cite{zhangContrastiveSelfSupervisedLearning2021} and MoCo \cite{xiaSelfSupervisedTextIndependentSpeaker2021} have been successfully applied to this field of research by focusing on how to define the model architecture and sample negative pairs for the training. These approaches rely on objective functions based on the Normalized Temperature-scaled Cross Entropy (\textit{NT-Xent}) loss. According to the self-supervised assumption, the contrastive loss aims at reducing the distance between positive pairs while increasing the distance between negative pairs in a latent space. 

Additive Margin from CosFace (\textit{AM-Softmax}) \cite{cosface} and Additive Angular Margin from ArcFace (\textit{AAM-Softmax}) \cite{arcface} have been successfully applied to improve angular softmax-based objective functions which are at the core of supervised SR systems \cite{utterancelevelaggforsv, largemarginsoftmaxforsv, 8683649}. Inspired by the performance obtained with these techniques on speaker verification, we introduce \textit{additive margin} into the \textit{NT-Xent} loss to improve the discriminative capacity of the embeddings by increasing speaker separability. We demonstrate the importance of this concept in a self-supervised setup as we obtained better performance when training SimCLR and MoCo SSL frameworks with \textit{additive margin}. We also verify the impact of class collisions as \textit{additive margin} could be applied between two samples from the same class. Moreover, we show that using a \textit{symmetric} contrastive loss using all possible positive and negative pairs is essential to improve the downstream performance of SimCLR.

This work extends the study presented in \cite{lepageAdditiveMargins2023} and provides results with models trained on VoxCeleb2 rather than VoxCeleb1. With this larger training set, we have seen that the projector is not useful. Moreover, we implement \textit{additive margin} in the MoCo framework to extend this concept to other SSL methods. Finally, we verify that class collisions are not an issue when implementing \textit{additive margin} in a self-supervised training scenario.

Our self-supervised training framework is described in Section~\ref{sec:method}. In particular, we revisit the standard contrastive loss and introduce \textit{additive margin} by defining \textit{NT-Xent-AM}. In Section~\ref{sec:experimental_setup}, we present our experimental setup. We report our results and assess the effect of \textit{additive margin} in Section~\ref{sec:results}. Finally, we conclude in Section~\ref{sec:conclusion}.
\section{Method}
\label{sec:method}

Our self-supervised training framework relies on a simple siamese architecture to produce a pair of embeddings for a given unlabeled utterance.

For each training step, we randomly sample $N$ utterances from the dataset. Let $i \in I \equiv\{1 \ldots N\}$ be the index of a random sample from the mini-batch.  We extract two non-overlapping frames, denoted as $\boldsymbol{x}_i$ and $\boldsymbol{x}_i^{\prime}$, from each utterance. Then, we apply random augmentations to both copies and use their mel-scaled spectrogram as input features. Using distinct frames and applying data augmentation is fundamental to avoid collapse and to produce robust representations that mainly contain speaker identity. An encoder transforms $\boldsymbol{x}_i$ and $\boldsymbol{x}_i^{\prime}$ to their respective representations $\boldsymbol{z}_i$ and $\boldsymbol{z}_i^{\prime}$. During training, mini-batches are created by stacking $\boldsymbol{z}_i$ samples into $\boldsymbol{Z}$ and $\boldsymbol{z}_i^{\prime}$ samples into $\boldsymbol{Z'}$. Representations are used to perform speaker verification and to compute the loss.

\subsection{Contrastive-based self-supervised learning}

Contrastive learning aims at maximizing the similarity within positive pairs while maximizing the distance between negative pairs. In self-supervised learning, supervision is provided by assuming that two randomly sampled utterances belong to different speakers. Positive pairs are constructed with embeddings derived from the same utterances, while negative pairs are sampled from the current mini-batch or a memory queue.

We start by defining the similarity between two embeddings $\boldsymbol{u}$ and $\boldsymbol{v}$ as $\ell(\boldsymbol{u}, \boldsymbol{v})=e^{\cos\left(\theta_{\boldsymbol{u}, \boldsymbol{v}}\right) / \tau}$ where $\tau$ is a temperature scaling hyper-parameter and $\theta_{\boldsymbol{u}, \boldsymbol{v}}$ is the angle between the two vectors $\boldsymbol{u}$ and $\boldsymbol{v}$. $\cos(\theta_{\boldsymbol{u}, \boldsymbol{v}})$ corresponds to the cosine similarity and is obtained by computing the dot product between the two $l_2$ normalized embeddings $\boldsymbol{u}$ and $\boldsymbol{v}$. Then, the Normalized Temperature-scaled Cross Entropy loss (\textit{NT-Xent}) is defined as

\begin{equation}
    \mathcal{L}_{\text {NT-Xent}}=- \frac{1}{N} \sum_{i \in I} \log \frac{\ell \left(\boldsymbol{z}_i, \boldsymbol{z}_{i}^{\prime} \right)}{\sum\limits_{a \in I} \ell \left(\boldsymbol{z}_i, \boldsymbol{z}_{a}^{\prime} \right)}. \label{eq:eqn1}
\end{equation}

We refer to $\boldsymbol{z}_i$ as the \textit{anchor}, $\boldsymbol{z}_i^{\prime}$ as the \textit{positive}, and $\boldsymbol{z}_a^{\prime}$ as a \textit{negative} when $a \neq i$. Thus, $N$ positive pairs are created, and compared to $N-1$ negatives.

Finally, we propose to compute the similarity of positive and negative pairs differently to introduce \textit{additive margin} later on, such that
\begin{equation}
    \mathcal{L}_{\text {NT-Xent}}=\\- \frac{1}{N} \sum_{i \in I} \log \frac{\ell^{+} \left(\boldsymbol{z}_i, \boldsymbol{z}_i^{\prime}\right)}{\ell^{+} \left(\boldsymbol{z}_{i} , \boldsymbol{z}_i^{\prime} \right) + \mkern-6mu\sum\limits_{a \in A(i)}\mkern-6mu \ell^{-} \left(\boldsymbol{z}_{i} , \boldsymbol{z}_a^{\prime} \right)},
\end{equation}
where $\ell^{+}(\mathbf{u}, \mathbf{v}) = \ell^{-}(\mathbf{u}, \mathbf{v}) = e^{\cos \left(\theta_{\boldsymbol{u}, \boldsymbol{v}}\right) / \tau}$ and $A(i) \equiv I \setminus \{i\}$.

\subsubsection{SimCLR -- Sampling negatives from the mini-batch}

Based on the above-mentioned technique, we adopt the SimCLR training framework, depicted in Figure~\ref{fig:training_framework} (a). The previous formulation of the contrastive loss, which has been adopted in \cite{huhAugmentationAdversarialTraining2020, zhangContrastiveSelfSupervisedLearning2021}, does not consider all possible positive and negative pairs. Instead, following the original implementation \cite{chenSimpleFrameworkContrastive2020}, we propose to use the \textit{symmetric} formulation of the \textit{NT-Xent} loss to increase the number of contrastive samples.

We now consider $\boldsymbol{z}_i$ to be the $i$-th element of a set of all embeddings created by concatenating $\boldsymbol{Z}$ and $\boldsymbol{Z}^{\prime}$. Let $i \in \hat{I} \equiv\{1 \ldots 2 N\}$ be the index of a random augmented frame, $j(i)$ be the index of the other augmented frame from the same utterance being in the same mini-batch and $\hat{A}(i) \equiv \hat{I} \setminus \{i, j(i)\}$. The \textit{Symmetric NT-Xent} loss is defined as
\begin{equation}
    \textstyle
    \mathcal{L}_{\text {NT-Xent}}^{(S)}=- \frac{1}{2N} \sum_{i \in \hat{I}} \log \frac{\ell^{+} \left(\boldsymbol{z}_i, \boldsymbol{z}_{j(i)}\right)}{\ell^{+} \left(\boldsymbol{z}_i, \boldsymbol{z}_{j(i)}\right) + \sum\limits_{a \in \hat{A}(i)} \ell^{-} \left(\boldsymbol{z}_i, \boldsymbol{z}_a \right)}
\end{equation}

Note that this results in $2N$ positive pairs which will be compared against a set of $2(N-1)$ negatives (all utterances in the mini-batch except the positive and the anchor).

\subsubsection{MoCo -- Sampling negatives from a memory queue}

We also explore the MoCo training framework, depicted in Figure~\ref{fig:training_framework} (b). Instead of being constrained by the batch size $N$, this method samples negatives from a larger memory queue to increase the number of negative samples. The major difference with SimCLR is that the gradient is not propagated through the second branch of the siamese architecture as the \textit{key} (\textit{momentum)} encoder is frozen and updated with an Exponential Moving Average (EMA) of the \textit{query} encoder parameters. The queue contains $K$ elements and is updated dynamically at each step of the training with embeddings from the \textit{key} encoder. Thus, all $N$ positive pairs are compared to $K$ negatives.

We define the \textit{Queue-based NT-Xent} loss such that
\begin{equation}
    \mathcal{L}_{\text {NT-Xent}}^{(Q)}=- \frac{1}{N} \\ \sum_{i \in I} \log \frac{\ell^{+} \left(\boldsymbol{z}_i, \boldsymbol{z}_i^{\prime}\right)}{\ell^{+} \left(\boldsymbol{z}_{i} , \boldsymbol{z}_i^{\prime} \right) + \sum\limits_{b \in B} \ell^{-} \left(\boldsymbol{z}_{i} , \boldsymbol{q}_b \right)},
\end{equation}
where $B \equiv\{1 \ldots K\}$ and $\boldsymbol{q}_i$ is the $i$-th element of the queue.


\subsection{Improving speaker separability with additive margin}

The contrastive loss is at the core of most self-supervised learning frameworks. However, it aims to penalize classification errors instead of producing discriminative representations relevant to the context of speaker verification.

Inspired by state-of-the-art techniques for face recognition, \textit{additive margin} were successfully applied for training end-to-end speaker verification models in a supervised way \cite{utterancelevelaggforsv, largemarginsoftmaxforsv, 8683649}, which justifies our motivation to adapt this concept for self-supervised learning. Following CosFace (AM-Softmax) \cite{cosface}, we introduce \textit{additive margin} in cosine space to increase the similarity of same-speaker embeddings and improve the discriminative capacity of our contrastive-based objective functions.

We define the Normalized Temperature-scaled Cross Entropy with Additive Margin loss (\textit{NT-Xent-AM}), $\mathcal{L}_{\text{NT-Xent-AM}}$, by setting 
\begin{equation}
    \ell^{+}(\mathbf{u}, \mathbf{v})=e^{\left(\cos\left(\theta_{\mathbf{u}, \mathbf{v}}\right)-m\right) / \tau},
\end{equation}
where $m \geq 0$ is a fixed scalar introduced to control the magnitude of the cosine margin, while $\ell^{-}(\mathbf{u}, \mathbf{v})$ remains unchanged.

Considering an \textit{anchor} $\boldsymbol{z}_a$, a \textit{positive} $\boldsymbol{z}_p$, and a \textit{negative} $\boldsymbol{z}_n$, the effect of this technique is to learn a constraint of the form $\cos \left(\theta_{\boldsymbol{z}_a, \boldsymbol{z}_p}\right) - m > \cos \left(\theta_{\boldsymbol{z}_a, \boldsymbol{z}_n}\right)$ instead of $\cos \left(\theta_{\boldsymbol{z}_a, \boldsymbol{z}_p}\right) > \cos \left(\theta_{\boldsymbol{z}_a, \boldsymbol{z}_n}\right)$, which set the positive similarity at least greater than the maximal negative similarity plus the margin constant.
\section{Experimental setup}
\label{sec:experimental_setup}

\subsection{Datasets and feature extraction}
\label{subsec:datafeat}
Our method is trained on the VoxCeleb2 dev set which is composed of $1,092,009$ utterances from $5,994$ speakers. The evaluation is performed on VoxCeleb1 \textit{original} test set containing $4,874$ utterances from $40$ speakers. According to our self-supervised training protocol, speaker labels are discarded.

The duration of extracted audio segments is $2$ seconds during training and $3.5$ seconds when performing speaker verification. Input features are obtained by computing $40$-dimensional log-mel spectrogram features with a Hamming window of $25$ ms length and $10$ ms frame-shift. As training data consists mostly of continuous speech segments, Voice Activity Detection (VAD) is not applied. The input features are normalized using instance normalization.

\subsection{Data-augmentation}
\label{subsec:dataaug}
Self-supervised learning frameworks commonly rely on extensive data-augmentation techniques to learn representations robust against extrinsic variabilities (noise from the environment, mismatching recording device...). Because positives are sampled from the same utterance, providing different augmented versions of the same utterance is fundamental to avoid encoding channel characteristics, allowing speaker identity to be the only distinguishing factor between two representations.

Thus, at each training step, we randomly apply two types of transformations to the input signal. First, we add background noises, overlapping music tracks, or speech segments using the MUSAN corpus. To simulate various real-world scenarios, we randomly sample the Signal-to-Noise Ratio (SNR) between $\left[13; 20\right]$ dB for speech, $\left[5; 15\right]$ dB for music, and $\left[0; 15\right]$ dB for noises. Then, to further enhance the robustness of our model, we apply reverberation to the augmented utterances using the simulated Room Impulse Response database.

\subsection{Models architecture and training}
\label{subsec:model}
The encoder neural network follows the Fast ResNet-34 \cite{chungDefenceMetricLearning2020} model architecture with $512$ output units. Utterance-level representations are generated using Self-Attentive Pooling (SAP).

Unlike most self-supervised frameworks we do not rely on a projector module, similar to other works on SSL for SV \cite{zhangContrastiveSelfSupervisedLearning2021, huhAugmentationAdversarialTraining2020}. Our previous work \cite{lepageLabelEfficientSelfSupervisedSpeaker2022} relied on a projector as it resulted in better performance when training on VoxCeleb1. As positive pairs are extracted from the same utterances, contrastive SSL methods are prone to encoding too much channel information while the projector allows extracting latent representations containing more speaker information. Thus, we hypothesize that this module was required on a smaller training set (VoxCeleb1) to reduce the overfitting on channel characteristics, avoiding a misalignment between the SSL objective and the downstream SV task.

We set the temperature to $\tau=\frac{1}{30}$ to match the scaling factor $s=30$ used commonly when training AM-Softmax supervised losses for speaker verification. When training the framework based on MoCo, we use a memory queue of $10,000$ negatives and update the frozen encoder with an exponential moving average (EMA) using a coefficient of $0.999$. We optimize the model with Adam optimizer and a learning rate of $0.001$, which is reduced by $5\%$ every $5$ epochs, with no weight decay. We use a batch size of $200$ and train SimCLR for $150$ epochs and MoCo for $100$ epochs. Our implementation is based on the PyTorch framework and we conduct our experiments on 2 $\times$ NVIDIA Tesla V100 16 GB.

\subsection{Evaluation protocol}
\label{subsec:evalprotocol}
To evaluate our model's performance on speaker verification, we extract embeddings from $10$ evenly spaced frames for each test utterance. Then, to determine the scoring of each trial, we compute the average of the cosine similarity between all $100$ combinations of $l_2$-normalized embeddings pairs. Following VoxCeleb and NIST Speaker Recognition evaluation protocols, we report the performance of our model in terms of Equal Error Rate (EER) and minimum Detection Cost Function (minDCF) with $P_{target} = 0.01$, $C_{\text{miss}}=1$ and $C_{\text{fa}}=1$.
\newcommand{\xmark}{\scalebox{0.85}{\usym{2613}}}

\section{Results and discussions}
\label{sec:results}

\subsection{Effect of our improvements to the self-supervised contrastive loss}

We conduct a study to assess the role of the different components of our self-supervised training framework based on SimCLR and report the results in Table~\ref{tab:improvements}. Our baseline, similar to the work of \cite{zhangContrastiveSelfSupervisedLearning2021}, trained with the $\mathcal{L}_{\text{NT-Xent}}$ loss achieves $8.98\%$ EER and $0.6714$ minDCF.

\begin{table}[t]
  \caption{The effect of the symmetric contrastive loss and additive margin on SimCLR self-supervised training for speaker verification.\\}
  \label{tab:improvements}
  \centering
  \begin{tabular}{lcS[table-format=1.2]cc}
    \toprule
    \textbf{Loss} & \textbf{Sym.} & \textbf{Margin} & \textbf{EER (\%)}     & \textbf{$\text{minDCF}_\text{0.01}$} \\
    \midrule
    $\mathcal{L}_{\text{NT-Xent}}$     & \xmark  & 0   &  $8.98$ & $0.6714$ \\
    \midrule
    $\mathcal{L}_{\text{NT-Xent}}^{(S)}$    & \checkmark & 0   & $8.41$ & $0.6235$ \\
    \midrule
    \multirow{3}{*}{$\mathcal{L}_{\text{NT-Xent-AM}}^{(S)}$} & \multirow{3}{*}{\checkmark} & 0.05 & $8.35$ & $0.6098$  \\
    ~ & ~ & 0.1 &  $7.85$ & $0.6168$ \\
    ~ & ~ & 0.2 &  $8.13$ & $0.6211$ \\
    \bottomrule
  \end{tabular}
\end{table}

\begin{table}[t]
  \caption{The effect of additive margin on MoCo self-supervised training for speaker verification.\\}
  \label{tab:moco}
  \centering
  \begin{tabular}{lS[table-format=1.1]cc}
    \toprule
    \textbf{Loss} & \textbf{Margin} & \textbf{EER (\%)} & \textbf{$\text{minDCF}_\text{0.01}$} \\
    \midrule
    $\mathcal{L}_{\text{NT-Xent}}^{(Q)}$    & 0   & $9.59$ & $0.6974$ \\
    \midrule
    $\mathcal{L}_{\text{NT-Xent-AM}}^{(Q)}$ & 0.1 & $9.36$ & $0.6403$ \\
    \bottomrule
  \end{tabular}
\end{table}

Firstly, we show that relying on more positive and negative pairs with the symmetric contrastive loss results in $8.41\%$ EER and $0.6235$ minDCF. This validates our intuition that providing more supervision and more negative samples is beneficial to improve the self-supervised system's downstream performance.

Secondly, we evaluate the role of \textit{additive margin} by showing that the choice of the margin value significantly impacts speaker verification results. The best setting is obtained with $m=0.1$ achieving $7.85\%$ EER and $0.6168$ minDCF, representing $12.6\%$ relative improvement of the EER over the baseline. Intuitively, a small margin does not affect the results, but a very large margin increases the training task complexity. It is noteworthy that $m=0.1$ corresponds to the value often used for AM-Softmax supervised training and that, according to our previous experiments on VoxCeleb1 \cite{lepageAdditiveMargins2023}, learning the margin value jointly with the model degrades the performance.

Finally, we observed the same positive effect when training MoCo with \textit{additive margin}. As shown in Table~\ref{tab:moco}, $\mathcal{L}_{\text{NT-Xent-AM}}^{(Q)}$ with $m=0.1$ improves the EER from $9.59\%$ to $9.36\%$ when compared to $\mathcal{L}_{\text{NT-Xent}}^{(Q)}$. The performances on MoCo are not as competitive compared to SimCLR because we did not optimize the different hyper-parameters related to this algorithm.

Thus, \textit{additive margin} is essential in self-supervised contrastive frameworks to improve results in speaker verification. We hypothesize that these improvements could benefit other downstream tasks related to verification.

\subsection{Impact of class collisions and class imbalance on the self-supervised training}

Class collisions occur when at least one negative, sampled from the mini-batch or the memory queue, belongs to the same speaker as the positive sample. The resulting false negative samples could represent an issue when using \textit{additive margin} with contrastive self-supervised training.

We conducted two experiments to test the effect of class collisions and class imbalance on the training. For the first experiment, we trained our SimCLR framework but used the train set labels to prevent class collisions in each mini-batch. For the second experiment, we limited the number of samples for over-represented speakers in the training set.

As shown in Table~\ref{tab:impact_ssl}, removing class collisions does not result in better downstream performance. Our assumption is that the probability of class collisions is too small as VoxCeleb2 contains many speakers compared to the batch size. Moreover, preventing class imbalance by limiting the number of samples for some speakers does not improve final speaker verification results. Thus, \textit{additive margin} can be used reliably in a self-supervised scenario.

\begin{table}[t]
  \caption{The impact of class collisions and class imbalance, which stems from the self-supervised training, on speaker verification results. The baseline is our SimCLR framework trained with $\mathcal{L}_{\text{NT-Xent-AM}}^{(S)}$ ($m=0.1$).\\}
  \label{tab:impact_ssl}
  \centering
  \begin{tabular}{ccccc}
    \toprule
    \makecell{\textbf{Class}\\ \textbf{collisions}} & \makecell{\textbf{Class}\\ \textbf{imbalance}} & \textbf{EER (\%)}     & \textbf{$\text{minDCF}_\text{0.01}$} \\
    \midrule
    \checkmark & \checkmark & $7.85$ & $0.6168$ \\
    \midrule
    \xmark & \checkmark & $7.95$ & $0.6241$ \\
    \xmark & \xmark & $8.41$ & $0.6390$ \\
    \bottomrule
  \end{tabular}
\end{table}

\subsection{Visualization of trials score distribution with and without additive margin}

\begin{figure}[t]
  \centering
  \includegraphics[width=\linewidth]{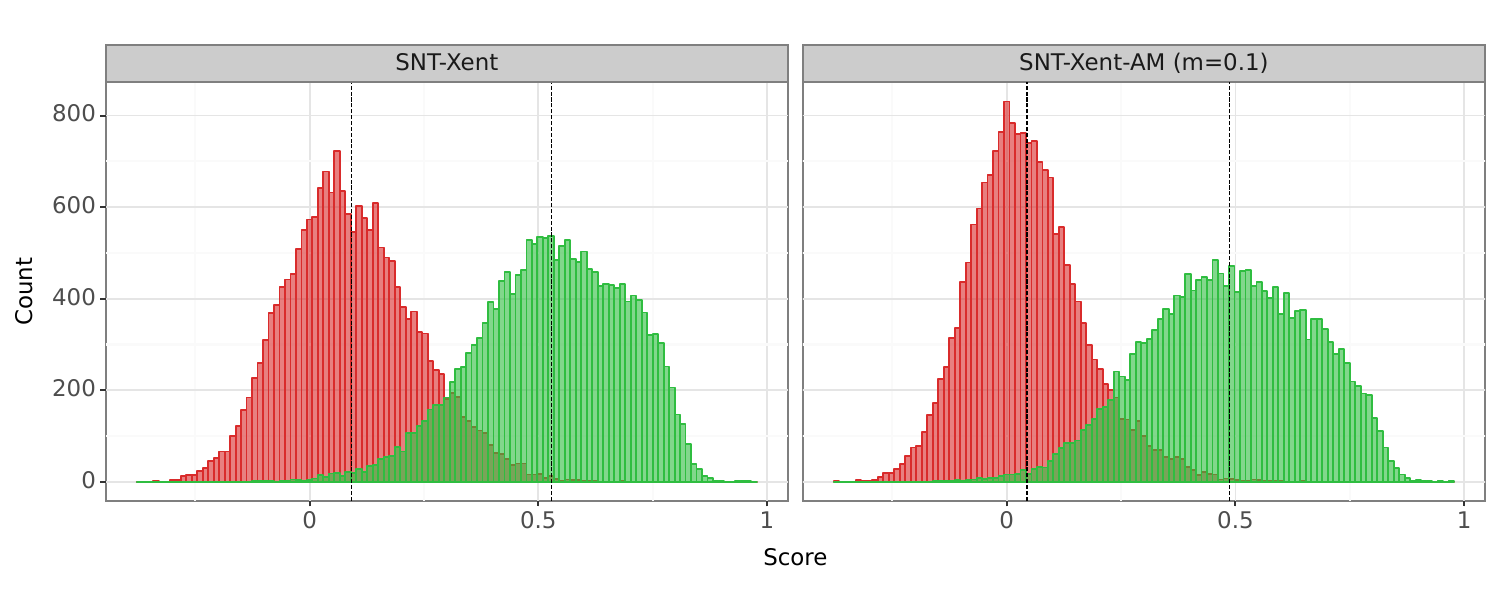}
  \caption{Positive (green) and negative (red) trials score distribution on VoxCeleb1-O and obtained after training our SimCLR framework with $\mathcal{L}_{\text{NT-Xent}}^{(S)}$ and $\mathcal{L}_{\text{NT-Xent-AM}}^{(S)}$ ($m=0.1$) losses. The mean of each distribution is represented by a dashed line.}
  \label{fig:score_dist}
\end{figure}

In Figure~\ref{fig:score_dist}, we plot the distribution of scores computed on VoxCeleb1 original test set to assess the effect of \textit{additive margin} on SimCLR with $m=0.1$. Visually, we can notice two improvements when introducing \textit{additive margin}: (1) the means of the two distributions are further apart; (2) the spread of negative scores is smaller.

This observation is consistent with the improvement of the EER and the minDCF. It shows that \textit{additive margin} positively affects the discriminative capacity of representations learned by self-supervised systems designed for verification tasks.

\subsection{Comparison to other self-supervised contrastive methods for speaker verification}

\definecolor{Gray}{gray}{0.9}

\begin{table}[t]
  \small
  \caption{Final results of different self-supervised contrastive methods on speaker verification. Note that $\mathcal{L}_{\text{AP}}$ corresponds to $\mathcal{L}_{\text{NT-Xent}}$ without a temperature but a learnable weight and bias.\\}
  \label{tab:final_results}
  \centering
  \begin{tabular}{llcc}
    \toprule
    \textbf{Method} & \textbf{Loss} & \textbf{EER (\%)} & \textbf{$\text{minDCF}_\text{0.01}$} \\
    \midrule
    \rowcolor{Gray} AP \cite{huhAugmentationAdversarialTraining2020} & $\mathcal{L}_{\text{AP}}$ & $9.56$ & $-$ \\
    \rowcolor{Gray} SimCLR \cite{zhangContrastiveSelfSupervisedLearning2021} & $\mathcal{L}_{\text{AP}}$ & $8.28$ & $0.6100$  \\
    \rowcolor{Gray} MoCo \cite{xiaSelfSupervisedTextIndependentSpeaker2021} & $\mathcal{L}_{\text{NT-Xent}}$ & $8.23$ & $0.5900$ \\
    \midrule
    SimCLR & $\mathcal{L}_{\text{NT-Xent-AM}}^{(S)}$ & $7.85$ & $0.6168$ \\
    MoCo   & $\mathcal{L}_{\text{NT-Xent-AM}}^{(Q)}$ & $9.36$ & $0.6403$ \\
    \bottomrule
  \end{tabular}
\end{table}

We report the final results on speaker verification in Table~\ref{tab:final_results} to compare to other contrastive self-supervised methods. Methods reported in the top part are trained with objective functions equivalent to $\mathcal{L}_{\text{NT-Xent}}$. Thus, they do not rely on the symmetric formulation of the contrastive loss and do not compute the similarity of positive and negative pairs differently to introduce \textit{additive margin}.

We outperform other contrastive approaches as the best result is obtained with our framework based on SimCLR, which achieves $7.85\%$ EER. Therefore, we showed that standard contrastive methods can be further improved by introducing simple changes tailored for the self-supervised training condition (providing more supervision with additional positives and negatives) and for the downstream task (increasing speaker separability for enhanced speaker verification).

\section{Conclusions}
\label{sec:conclusion}

This article introduces \textit{additive margin} to self-supervised contrastive frameworks by defining \textit{NT-Xent-AM} to learn more robust speaker representations. We demonstrated that this improvement results in a lower EER on the VoxCeleb test set and a better discrepancy between scores of positive and negative trials, even in a self-supervised training scenario. We showed that \textit{additive margin} combined with the symmetric formulation of the contrastive loss applied to SimCLR outperforms the other equivalent contrastive approaches. We expect NT-Xent-AM to be also effective in training SSL contrastive models for other tasks and modalities in image or speech processing (language recognition, ...).

\section{Acknowledgements}

This work was performed using HPC resources from GENCI-IDRIS (Grant 2023-AD011014623) and has been partially funded by the French National Research Agency (project APATE - ANR-22-CE39-0016-05).

\bibliographystyle{IEEEbib}
\bibliography{main}

\end{document}